\documentclass{ifacconf}

\usepackage{graphicx}      % include this line if your document contains figures
\usepackage{natbib}        % required for bibliography
\usepackage{amsmath,amssymb}
\usepackage{xcolor} 
\usepackage[table]{xcolor}  
\usepackage{comment}
\definecolor{darkgreen}{rgb}{0.0, 0.5, 0.0}
\usepackage{makecell}
\usepackage{algorithm}
\usepackage{algpseudocode}
\usepackage{times} % Litet fusk
\usepackage{subcaption}
\usepackage{bm}
\usepackage{microtype}
\usepackage{url}
\usepackage[backgroundcolor=teal!30]{todonotes}
\graphicspath{ {./Images/} } 

\renewcommand{\cite}{\citep}

\DeclareMathOperator*{\minimize}{minimize}

\DeclareMathOperator{\subjectto}{subject\ to}

\begin{document}
\begin{frontmatter}
	\title{Vessel Trajectory Prediction using COLREGs-aware Optimal Planning\thanksref{footnoteinfo}}
	% Title, preferably not more than 10 words.

	\thanks[footnoteinfo]{Conducted while D. Kaikkonen was an M.Sc student at Linköping University and performing his thesis work at ABB Corporate Research. \textcolor{red}{\copyright2026 IFAC the authors. This work has been accepted to IFAC for publication under a Creative Commons Licence CC-BY-NC-ND.}}

	\author[First]{David Kaikkonen},
	\author[ABB]{Fredrik Ljungberg},
	\author[LiU]{Erik Frisk}

	\address[First]{david.kaikkonen@nibe.se}
	\address[ABB]{ABB Corporate Research, Västerås, Sweden (fredrik.ljungberg@se.abb.com)}
	\address[LiU]{Department of Electrical Engineering, Linköping University, Sweden (erik.frisk@liu.se).}
	%\address[Second]{Colorado State University, 
	%   Fort Collins, CO 80523 USA (e-mail: author@lamar. colostate.edu)}
	%\address[Third]{Electrical Engineering Department, 
	%   Seoul National University, Seoul, Korea, (e-mail: author@snu.ac.kr)}

	\begin{abstract}                % Abstract of 50--100 words
		This paper presents a trajectory prediction method for marine vessels based on optimal planning.  Crude initial trajectories respecting static obstacles are first generated using A*-search to provide a feasible warm start. In the second step, a numerical optimizer is used to ensure COLREG compliance. The prediction problem is posed as sequential trajectory planning from the perspective of each surrounding vessel, requiring only their current positions, velocities, and intended destinations as input. As the latter is included in AIS messages, this enables faster predictions than learning-based methods that typically require longer data histories. The proposed method is validated using real-world scenarios constructed from AIS data.
	\end{abstract}


\end{frontmatter}
%===============================================================================

\section{Introduction}
Autonomous vehicles are becoming increasingly common, and research is being conducted across several areas of transportation, including the maritime sector. A key step in automated control of marine vessels is determining a dynamically feasible path, and since the explored navigation area is sometimes occupied by other maritime traffic, this planned path must be updated in real-time based on sensor feedback. Further, beyond being collision-free, the path must adhere to the Convention on the International Regulations for Preventing Collisions at Sea (COLREGs), which govern how vessels should communicate and navigate during interactions to behave in a predictable and risk-averse manner \cite{maza2022colregs}.

Collision-free and COLREG-compliant paths can be obtained in several different ways \cite{ zhao2022intelligent, bergman2020optimization,zhang2022collision}, but all methods have in common that they are highly dependent on the availability of accurate information regarding the motions of the other vessels. Information about another vessel's current position, velocity, and heading direction can be obtained by fusing sensor data (e.g., from camera, LiDAR, or Radar) using a tracking algorithm. However, for automated maneuvering in highly trafficked areas, estimates of future trajectories are required that extend beyond the vessel's current state.

A naive way of predicting a vessel's future position is to simply assume that it will keep its heading and speed for the foreseeable future. Such a constant-velocity (CV) assumption can be valid in open-sea traffic where vessels mostly travel along straight paths, see e.g., \cite{zhang2024review,cademartori2023review}. However, in collision-avoidance scenarios, especially in highly trafficked areas like archipelagos where vessels frequently are forced to perform tight maneuvers, a CV~assumption can give  poor predictions unsuitable in collision avoidance scenarios.
% Despite this, many path planning approaches in the literature are based on CV assumptions \cite{zhang2024review,cademartori2023review}. %huang2020ship

There are also proposed techniques for predicting more advanced maneuvers where many are based on learning-based black-box methods \cite{li2024vessel,cheng2022}. Such approaches can give very accurate results, especially in scenarios for which there exists substantial amounts of training data. However, not being able to trace inevitable poor predictions back to physical principles is a significant limitation in safety-critical applications like marine navigation. Beyond the lack of interpretability, black-box methods struggle to generalize to new scenarios, such as uncommon traffic situations or deviating environmental conditions.

This paper explores an alternative method for predicting the motion of surrounding vessels by formulating the trajectory forecasting as an optimal control problem. The work is based on the results presented in \cite{johansson2025motion}. The main idea is to predict the motion of the other ships by sequentially planning a trajectory from each of their perspectives. More specifically, initial trajectories are generated using A*-search to provide a feasible warm start for a numerical optimizer, similarly to how paths are often planned for the ego vessel. This is readily possible for larger ships as they must submit information about their final destinations through the automatic identification system~(AIS). Beyond the goal point, the method relies only on a vessel’s current position and velocity, enabling faster predictions than black-box models that typically require long input histories. The approach incorporates basic physics and COLREGs and can be readily extended to account for environmental factors such as wind and waves, which is an additional advantage over learning-based methods. Validation in realistic simulation scenarios constructed from historical AIS data demonstrates its potential to capture plausible vessel behaviors.

\section{Problem and Method Outline}
% \section{Problem Description}
\label{Problem_Description}
%When relying on motion planning for predictions, the absence of a goal position is a factor of uncertainty. As there are two main parts of predicting vessel trajectories: Identifying the correct collision maneuver, and predicting the correct destination. The destination and goal point uncertainty can be alleviated by using AIS data, something that forms the basis of many vessel prediction algorithms. \textcolor{red}{[Källa till andra metoder vi nämnt i litteratur]}
The main purpose of the prediction method is to support motion planning. Safe trajectory planning in traffic scenarios requires accurate predictions with sufficient look-ahead to anticipate conflicts and assess risks posed by nearby vessels.
This creates a need to identify the vessels that are most critical to maneuver-safety, allowing for simpler prediction models for less critical vessels to improve computational efficiency.
The proposed method addresses this by combining an approximate long-term prediction for all vessels with a more detailed short-term prediction for those showing a high collision risk, subsequently referred to as target vessels.
The algorithm predicts the motion of one target vessel at a time and uses the cruder long-term prediction for the vessels currently not in focus. %To justify using A* search for the long-term prediction, it is assumed that all vessels intend to follow a planned path that emphasizes efficiency and fuel consumption.

The long-term prediction estimates the vessel's complete trajectory, considering only static obstacles such as islands and shallow water. In contrast, the short-term prediction is continuously updated to assess collision risks and identify what COLREGs maneuver the vessel is expected to take to mitigate them. The long-term prediction serves as input for refining the short-term prediction by narrowing down the search space for reasonable paths and additionally provides approximate trajectories of other surrounding vessels.

The following sections outline the main steps of the proposed prediction algorithm, and a summary in pseudocode is presented in Algorithm~\ref{alg:cap}.

%\textbf{Stage 1: Initializing Algorithm}
%1%.1 Identify Surrounding vessels \\
%1.2 Designate target vessel \\
%1.3 Set prediction parameters (Safety distance, COLREG's identification distance, vessel max velocity \& turn-rate)
%\textbf{Stage 2: Long-term A*}\\
%2.1: Create node network for A*, \\
%2.2: Remove nodes inside static obstacles, make nodes nearby obstacles more expensive\\
%2.3: Pull final destination for vessels from AIS-data or goal-point estimation\\
%2.4: Obstacle avoidant long-term A* for all vessels\\
%2.5: Estimate CPA between target vessel and surrounding vessels \\
%\textbf{Stage 3: Short-term A*, warm-start}\\
%3.1 Estimate short-term goal point along long-term A*-path\\
%3.2 Identify COLREGs, create forbidden zones\\
%3.3 Short term A* solution to warm-start OCP with initialization of optimization variables\\
%\textbf{Stage 4: Optimal Control Problem}\\
%4.1 Solve optimal control problem to generate prediction for the target vessels' trajectory
%4.2 Save solution and vessel data for the next iteration of the algorithm and displaying results
% \begin{algorithm}[h]
% \caption{Vessel Prediction (Simplified)}\label{alg:simple}
% \begin{algorithmic}[1]
% \State \textbf{Stage 1: Initialization} – Identify vessels, set parameters
% \State \textbf{Stage 2: Long-term Planning} – Compute A* paths avoiding static obstacles, estimate CPA
% \State \textbf{Stage 3: Short-term Planning} – Warm-start A* toward short-term goal, consider COLREGs
% \State \textbf{Stage 4: Optimal Control} – Solve OCP using warm-start and constraints to predict trajectory
% \end{algorithmic}
% \end{algorithm}
\begin{algorithm}[h]
	\caption{Vessel prediction}\label{alg:cap}
	\begin{algorithmic}[1]

		\State \textbf{Stage 1: Initializing Algorithm}
		\State Identify surrounding vessels \& designate target vessel
		\State Set prediction parameters (safety distance, COLREGs distance, max velocity, turn-rate),

		\State \textbf{Stage 2: Long-term A*}
		\State Create node network for A*, remove nodes inside static obstacles, increase cost near obstacles
		\State Retrieve final destination from AIS data or goal estimation for all vessels
		\State Compute long-term A* path for all vessels, which avoids static obstacles,
		\State Estimate CPA between target vessel and surrounding vessels,

		\State \textbf{Stage 3: Short-term A* Warm-start}
		\State Estimate short-term goal point along long-term A* path for target vessel,
		\State Identify COLREGs and forbidden zones at CPA between target vessel and surrounding vessels,
		\State Solve A* towards short-term goal point, calculate control parameters used,

		\State \textbf{Stage 4: Optimal Control Problem}
		\State Use CPA, COLREGs \& forbidden zones as inputs to the OCP,
		\State Use short-term A* solution and calculated control parameters as warm start for OCP,
		\State Define vessel kinematic model and set cost function, penalizing travel time, steering inputs and acceleration,
		\State Solve OCP to generate trajectory prediction for target vessel,
		\State Save prediction and vessel data for the next iteration of the prediction algorithm and visualization.

	\end{algorithmic}
\end{algorithm}
%This is a multi-vessel prediction model, where one target vessel is chosen to extract a detailed prediction of their future trajectory using both the long- and short-term part of the model. The simpler long-term prediction is applied to all surrounding vessels to serve as input for refining the prediction of the target vessel.
%In the context of maritime navigation, motion planners need to apply a certain amount of caution to account for the inherently slow dynamics of vessels, and the strict safety standards
\subsection{Long-Term Prediction}
\begin{figure}[htbp]
	\centering
	\includegraphics[width=0.44\linewidth]{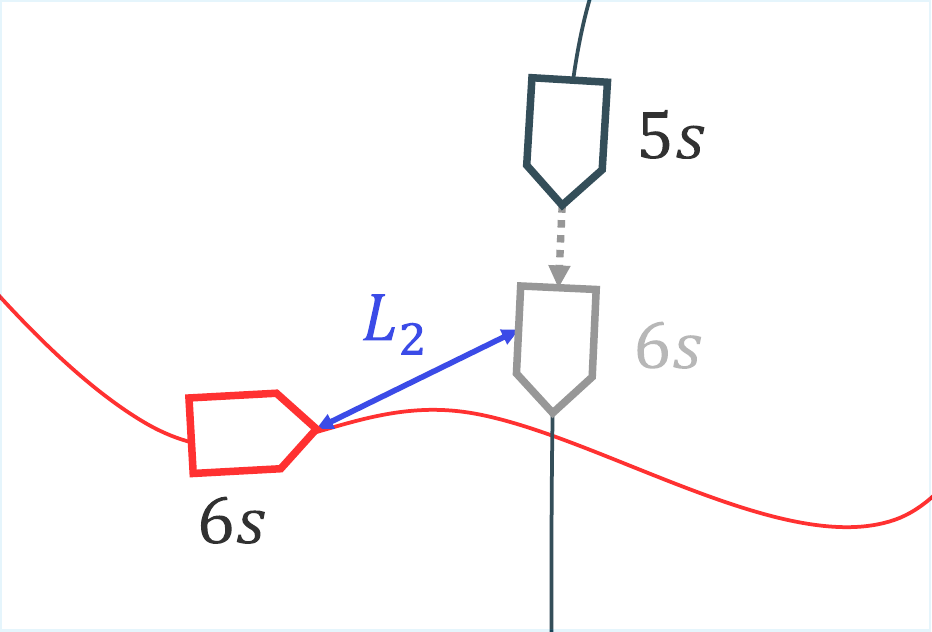}
	\caption{CPA estimate where L2 = distance at CPA.}
	\label{fig:node2_time6s}
\end{figure}

The long-term prediction outlined in Stage 2 of Algorithm~\ref{alg:cap} is a large-scale trajectory, providing a general idea of how the vessel will navigate in an environment without dynamic obstacles. It is assumed that vessels plan their path to minimize travel time and fuel consumption. As such, A* search is chosen as an efficient method to find an optimal path where islands and shallow waters are treated as obstacles, and the destination is supplied by AIS data.
%%It uses A* search, where islands and shallow water are treated as obstacles, and takes the destination supplied by AIS as the goal point. -- Vad som stod innan IFAC-Kommentarsändring, det utkommenterat under är från en äldre revidering
%If AIS data is unavailable, the prediction uses a scanning range for potential destinations. %and a past-trajectory check to identify if a vessel is turning to adjust its course towards its destination, or avoid an obstacle.% as shown in Fig.~\ref{fig:factors}.
\begin{comment}
\begin{figure}[h]
	\centering
	\begin{minipage}[b]{0.36\linewidth}
		\centering
		\includegraphics[width=\linewidth,height=0.3\textheight,keepaspectratio]{scanning_ny_1.pdf}
	\end{minipage}
	\hfill
	\begin{minipage}[b]{0.42\linewidth}
		\centering
		\includegraphics[width=\linewidth,height=0.3\textheight,keepaspectratio]{2stepback.pdf}
	\end{minipage}
	\caption{Scanning range (left) \& past-trajectory check (right) for the long-term prediction.}
	\label{fig:factors}
\end{figure}
\end{comment}
This prediction is applied to the target vessel and all nearby vessels, and the predicted trajectories are used in step 8 of Algorithm~\ref{alg:cap} to estimate the closest points of approach (CPA) between them. As shown in Fig.~\ref{fig:node2_time6s}, the CPA estimation evaluates the positions of the vessels relative to each other at every second along their predicted trajectories. This is used to identify when the target vessel is at the highest risk of collision with another vessel.

% [Kommenterade ut det under för det var en upprepning på det jag skriver i Short-term]
%A motion planner would typically use this information to calculate an efficient, collision-free path, with weights on factors such as collision risk and travel distance, depending on the model parameters. When using the same information for motion prediction, 

The method uses the CPA estimation to identify the point at which it believes a COLREGs maneuver will be taken. The focus on COLREGs provides a foundation for the prediction that is less subject to interpretation and has to be considered in any scenario, common or unfamiliar, and also adapts the prediction to the specific context of maritime navigation.
%These rules exist specifically to ensure predictable vessel behaviour, and would usually not be considered by a general optimization algorithm that only accounts for physical obstacles when determining navigable space.

\subsection{Short-Term Prediction}
\label{sec:short_term_prediction}

\begin{figure}[htbp]
	\centering
	\includegraphics[width=0.7\linewidth]{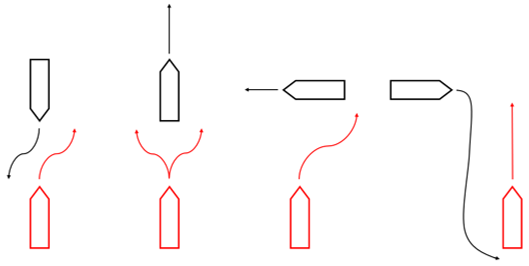}
	\caption{The right course of action in each COLREGs situation. Left to right: Head-On, Overtaking, Give-Way, Stand-On}
	\label{fig:colregs_sit}
\end{figure}

The short-term prediction outlined in Stage 3 of Algorithm~\ref{alg:cap} produces a detailed trajectory that respects both vessel dynamics and COLREGs, describing how the vessel is expected to behave in the near future.
It identifies traffic situations and generates forbidden zones to guide the prediction towards COLREGs compliance, and a short-term goal point chosen at a set distance along the long-term trajectory of the target vessel.
This is used to formulate the OCP outlined in Stage 4 of Algorithm~\ref{alg:cap}.
Unlike typical motion-planning objectives that optimize for criteria such as energy consumption or distance traveled, this prediction focuses on factors that strongly influence a vessel's choice of trajectory.
Consequently, the method emphasizes classifying interactions according to the traffic scenarios shown in Fig.~\ref{fig:colregs_sit}.
The forbidden zones, further detailed in Section~\ref{sec:colreg_zones}, ensure that the predicted trajectory does not violate COLREGs.

\subsection{System States and Time Discretization in OCP}

The level of detail in vessel dynamics and environmental factors have a high impact on the computational load of the algorithm. They must be carefully considered to be detailed enough to accurately capture vessel behavior while allowing the algorithm to frequently update the prediction as the situation evolves. As such, a simple vessel model detailing size, turn rate and maximum velocity defines the kinematic constraints for the OCP in step 16 of Algorithm~\ref{alg:cap}. With all inputs, the vessels $x$- and $y$-velocity, angular rate, $\dot{\theta}$, and acceleration, $a$, define a kinematic model, with a state vector containing the $x$- and $y$-position, orientation, $\theta$, and velocity, $v$:
% \vspace{-0.5em}
% \begin{equation} \label{eq:kinematic_model}
% 	f(x, u) = \begin{pmatrix} \dot{x} \\ \dot{y} \\ u_1 \\ u_2 \end{pmatrix}, \quad
% 	u_1 = \dot{\theta}, \quad u_2 = a.
% \end{equation}
\begin{equation}
	\label{eq:kinematic_model}
	\begin{aligned}
		\dot{x} & = v \cos(\theta), & \dot{\theta} & = u_1, \\
		\dot{y} & = v \sin(\theta), & \dot{v}      & = u_2. \\
	\end{aligned}
\end{equation}

% \begin{equation} \label{eq:state_vector}
% 	\begin{pmatrix}x & y & \theta & v\end{pmatrix}.
% \end{equation}
The final time, $T$, depends on the system inputs $u_i$ and is defined as an optimization variable with time-discretization
\vspace{-0.5em}
\begin{equation}
	dt = \frac{T}{N}
\end{equation}
with $N$ being the number of elements in the time-discretization, determining the size of each time step in the solution.
The time to travel the Euclidean distance between the start and goal points at maximum velocity is set as a minimum time constraint to ensure a physically feasible prediction, while the maximum for angular rate and acceleration are imposed as smoothness constraints.
The optimization problem is then defined as a nonlinear program,
\begin{equation}
	\label{eq:main_ocp}
	\begin{aligned}
		\minimize_{u_1, u_2} \quad & \alpha T + \beta u_1 + \gamma u_2      \\
		\subjectto \quad           & \text{Collision avoidance constraints} \\
		                           & \text{COLREG constraints}              \\
		                           & \text{Time constraints}                \\
		                           & \text{Kinematic constraints}           \\
		                           & \text{Smoothness constraints}
	\end{aligned}
\end{equation}
where $\alpha$, $\beta$, and $\gamma$ are weighted to scale each variable to comparable magnitudes.

%The prediction problem is divided into two scopes, each with their own difficulties: a long-term and a short-term prediction. Where a long term prediction is made to estimate the end position of the vessel, and multiple short-term predictions are conducted along the path to anticipate how the vessel will behave in scenarios where COLREGs need to be considered.

%As this paper mainly focuses on predicting trajectories on the basis of COLREGs interactions, secondary factors affecting the navigation environment such as wind, waves and currents are omitted. Vessel motion is also restricted to the horizontal plane, and simulations are conducted in daylight scenarios, where all surroundings obstacles are expected to be known and fully visible.

\subsection{COLREGs and Construction of Forbidden Zones}
\label{sec:colreg_zones}
%The algorithm is divided into a long- and short-term prediction. Where the long-term prediction estimates the full unobstructed trajectory of the vessel, considering only static obstacles. 

The construction of the input variables for the OCP is based on the method of identifying COLREGs scenarios by scanning areas separated by relative heading angles as described in \cite{maza2022colregs,breivik2017mpc}. The method builds upon calculating the interaction between the target vessel and each surrounding vessel one-by-one in order to evaluate if they have an impact on the outcome of the prediction. First, to determine if two vessels are approaching each other and may risk colliding, the dot product of the distance vector, $d_{rel}$, and relative velocity vector, $v_{rel}$
\[
	\bm{d}_{\text{rel}} \cdot \bm{v}_{\text{rel}} \propto \cos(\delta)
\]
is calculated. A negative scalar indicates that their distance is decreasing and a risk of collision cannot be ruled out. In such a scenario, COLREGs must be considered. The relative heading angle, $\theta_{COL}$ is computed from the target vessel's own heading, $\theta_{vessel}$, and the bearing to the obstacle vessel, $\theta_{bearing}$, as
%the target vessel's own heading, $\theta_{vessel}$, minus the bearing to the obstacle vessel, $\theta_{bearing}$, with the result wrapped between $\pm\pi$:
\[
	\theta_{COL}
	=
	\bigl(\theta_{vessel} - \theta_{bearing}\bigr)
	-2\pi \left\lfloor \frac{\theta_{vessel} - \theta_{bearing}}{2\pi} \right\rfloor,
\]

\[
	\theta_{COL} =
	\begin{cases}
		\theta_{COL},        & \theta_{COL} \le \pi, \\[6pt]
		\theta_{COL} - 2\pi, & \theta_{COL} > \pi.
	\end{cases}
\]

\begin{figure}[htbp]
	\centering
	\begin{minipage}[b]{0.48\linewidth}
		\centering
		\includegraphics[width=\linewidth,height=0.15\textheight,keepaspectratio]{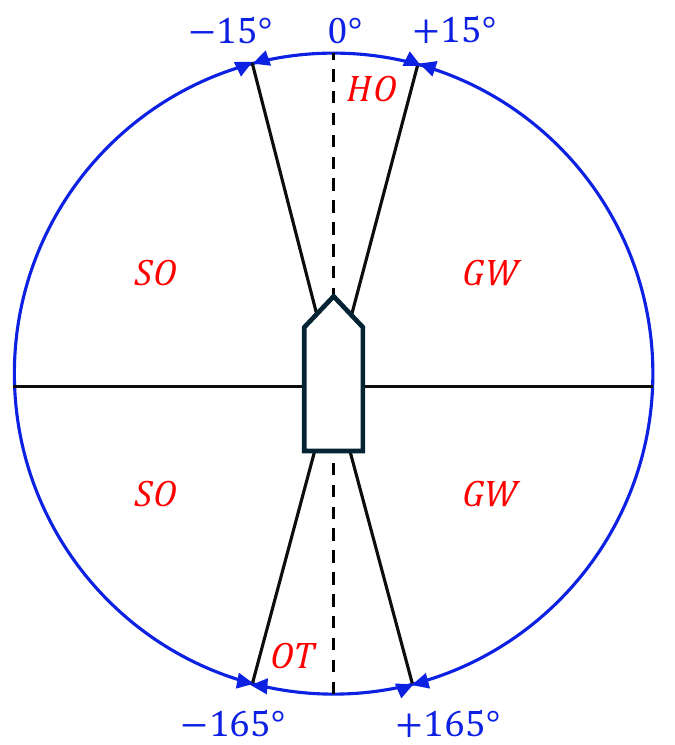}
		\phantomsubcaption
		\label{fig:angle}
	\end{minipage}
	\hfill
	\begin{minipage}[b]{0.48\linewidth}
		\centering
		\includegraphics[width=\linewidth,height=0.15\textheight,keepaspectratio]{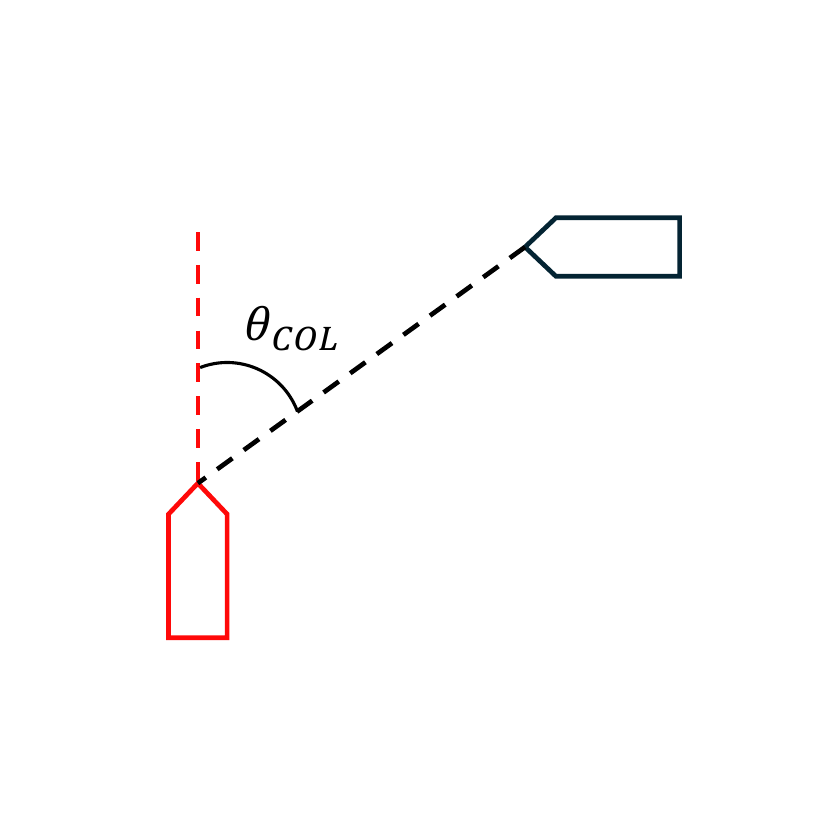}
		\phantomsubcaption
		\label{fig:relativeangle}
	\end{minipage}

	\caption{The identified COLREGs for each approach, with angles expressed in degrees for clarity.}
	\label{fig:headinganglecalculation}
\end{figure}

An example of this is shown in Fig.~\ref{fig:headinganglecalculation}.

\begin{figure}[htbp]
	\centering
	\includegraphics[width=0.78\linewidth]{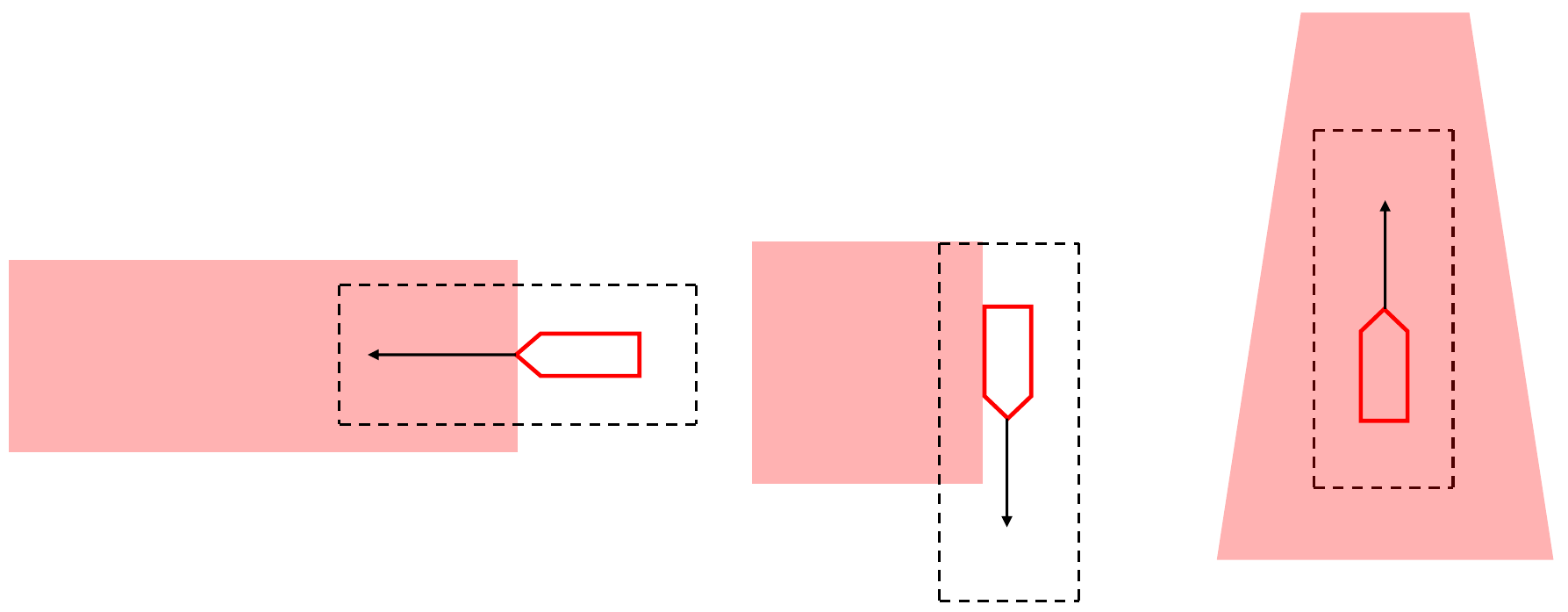}
	\caption{Forbidden zone creating a general safety margin around any vessel (dotted area), and forbidden zone for each specific COLREGs scenario (red area). From left to right: GW, HO, and OT.}
	\label{fig:forbidden}
\end{figure}

Depending on the identified COLREGs scenario, different forbidden zones are constructed around obstacles, which outline areas that imply a COLREGs violation if entered. These zones are generated in step 11 of Algorithm~\ref{alg:cap} and serve as constraints for the OCP-based short-term prediction. Examples are shown in Fig.~\ref{fig:forbidden}.

%There is no equivalence to these zones in a CV assumption, which often leads to larger areas than what is actually possible being considered feasible. This means that even in a scenario without static obstacles, a CV assumption fails to recognize traffic rules or surrounding vessels, and planning a path based upon such a prediction runs the risk of giving the planner a false sense of security, potentially leading to unsafe or COLREGs violating maneuvers.

%This means that even in an obstacle-free scenario, where a CV assumption performs better, planning a path based upon prediction results from a CV assumption still runs the risk of giving the planner a false sense of security.

\subsection{Collision Avoidance Constraints in OCP}
\label{distance_optimal}

Algorithm~\ref{alg:cap} in Section~\ref{Problem_Description}, includes a short-term prediction step where an Optimal Control Problem (OCP) is formulated to predict short-term trajectories for target vessels. A key aspect of this OCP is the inclusion of collision avoidance constraints to ensure that the predicted trajectories do not violate COLREGs by entering forbidden zones around other vessels. The collision avoidance constraints are here formulated using established techniques from convex optimization \cite{boyd2004convex}, which allow for efficient optimization and planning.

Optimization techniques for motion planning with obstacles are described in \cite{zhang2020optimization} where a dual formulation of distance constraints are used, which is also the approach briefly outlined below. See \cite{fan2024efficient} for alternative approaches and related discussions.

Each obstacle is represented as a convex polygon. Such a set can be described as an intersection of a number, $m$, of half-spaces, one for each face of the polygon. This means that the set, conveniently for optimization, can be represented by a linear matrix inequality in the form
\begin{equation*}
	A p \leq b
\end{equation*}
where $p\in\mathbb{R}^2$ is a point in the plane, the rows of $A\in\mathbb{R}^{m\times 2}$ are normals to each face in the obstacle polygon, and $b\in\mathbb{R}^m$ based of distances. Any $p$ satisfying the inequality is therefore inside the obstacle, and it is straightforward to transform the inequalities based on orientation and position of the obstacle.

A collision avoidance constraint in motion prediction corresponds to that the minimum distance between the vessel and an obstacle is larger than some minimum distance $d_{\text{min}}$.
A convex optimization formulation for the minimal distance from the EGO vessel position $p_0$ to an obstacle represented by a convex polygon can be expressed as
\begin{equation}
	\label{eq:primal_problem}
	\begin{aligned}
		\minimize_{p, d} \quad & \|d\|       \\
		\subjectto \quad       & A p \leq b  \\
		                       & p_0 - p = d
	\end{aligned}
\end{equation}
where $d$ is the distance to the obstacle. Now, the dual function $g(\lambda)$ for \eqref{eq:primal_problem}, where $\lambda$ are the dual variables, is a lower bound on the distance to the obstacle, and
\begin{equation*}
	g(\lambda) = (A p_0 - b)^T\lambda
\end{equation*}
for any $\lambda \geq 0$ satisfying $\|A^T\lambda\|\leq 1$ \cite[p. 399]{boyd2004convex}.
The dual function, with the constraints on $\lambda$, can therefore be used as a collision avoidance constraint in the OCP for motion planning \eqref{eq:main_ocp} for all time steps $k$ as
\begin{align*}
	(A_k p_k - b_k)^T\lambda_k & \geq d_\text{min} \\
	\|A_k^T\lambda_k\|         & \leq 1            \\
	\lambda_k                  & \geq 0
\end{align*}
where $A_k$ and $b_k$ is the polygon representation of the obstacle and $p_k$ and $\lambda_k$ are the position of the EGO vessel and dual variables at time step $k$ respectively.
% \begin{align*}
% 	\maximize_{\lambda} \quad & (Ax_0 - b)^T\lambda   \\
% 	\subjectto \quad          & \|A^T\lambda\| \leq 1 \\
% 	                          & \lambda \geq 0
% \end{align*}

Since the full OCP is inherently a non-convex optimization problem, the efficiency and feasibility of the optimization/planner will be sensitive to the optimizer's initialization. The following section thereby presents a basic procedure for initializing both the dual variables $\lambda$ and the predicted trajectories.

\subsection{Initialization of Optimization Variables}

\begin{figure}[htbp]
	\centering
	\includegraphics[width=0.45\linewidth]{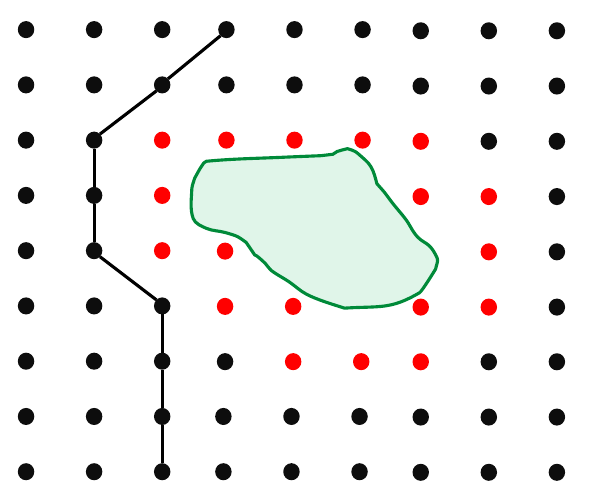}
	\caption{Initial path approximation using the A* algorithm, where red nodes are given a higher cost.}
	\label{fig:grannar2}
\end{figure}

In step 12 of Algorithm~\ref{alg:cap}, the A* algorithm is used to initialize the solver with an approximate solution as shown in Fig.~\ref{fig:grannar2}. In the A* graph-search, obstacle avoidance is ensured by removing nodes from the grid that are inside any obstacle, outlined in step 5 of Algorithm~\ref{alg:cap}.
% As this algorithm uses nodes when planning a path, the distance calculation in Section~\ref{distance_optimal} is not possible, or is at least needlessly accurate as only an approximate solution is being calculated. 
% Instead, A* is given obstacle avoidance by removing nodes from the grid which are inside any obstacle, outlined in step 5 of Algorithm~\ref{alg:cap}. 
As an additional safety margin, nodes which are nearby obstacles are identified by counting their amount of unobstructed neighbors. They are then given a higher cost, guiding the algorithm to choose a safer path if possible.
From this solution, the Cartesian position, heading angle and distance optimization variable, $\lambda$ is calculated as an approximate solution.

\begin{figure}[htbp]
	\centering
	\includegraphics[width=0.5\linewidth]{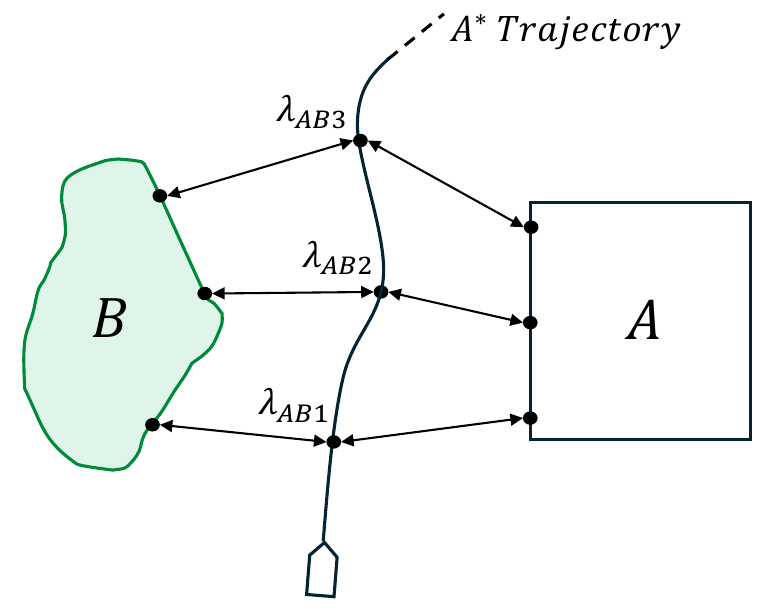}
	\caption{Calculated guess of $\lambda_{AB}$ from the approximated A* trajectory to each obstacle.}
	\label{fig:lambda_A*}
\end{figure}

The initial guess for the optimization variable $\lambda$ needs to consider the distance to every nearby obstacle in the scene. A method to find an approximate value of $\lambda$ for every obstacle is to pick nodes along the approximate A* path, which is guaranteed to not collide with any obstacle, and calculate $\lambda$ from those nodes to each obstacle in the scene using the minimization expressed in Section~\ref{distance_optimal}. This results in a final approximate solution shown in Fig.~\ref{fig:lambda_A*}

\section{Results}
The proposed prediction method is validated by analyzing AIS-data from real-world encounters using MarineTraffic \cite{MarineTraffic} and recreating them in a simulation environment to compare the algorithm's predicted path with the actual outcome. Optimization is performed using CasADi \cite{Andersson2019} using the solver IPOPT \cite{wachter2006implementation}. The Root Mean Square Error (RMSE) between the predicted and real path is also calculated for each prediction. The scenarios were not selected to highlight navigation errors, but rather because they provide realistic conditions suitable for validating the prediction algorithm.

\begin{comment}
The simulation environment is shown in Fig.~\ref{fig:potential_goals}. While the simulations feature encounters from different harbors, a map of the port of Helsinki is used to illustrate them as it was used in the creation of the algorithm in \cite{johansson2025motion}. Land and shallow waters are considered to be static obstacles, with a high- to guaranteed risk of grounding or collision. %Harbors, Beaches and Boat Berths are counted as potential destinations for vessels.

\begin{figure}[h]
	\centering
	\includegraphics[width=0.65\linewidth]{potential_goals.pdf}
	\caption{Helsinki Harbor in the simulation environment.}
	\label{fig:potential_goals}
\end{figure}
\end{comment}
\subsection{Double Head-On and Give-Way Predictions}

\begin{figure}[tbp]
	\centering
	\includegraphics[width=0.8\linewidth]{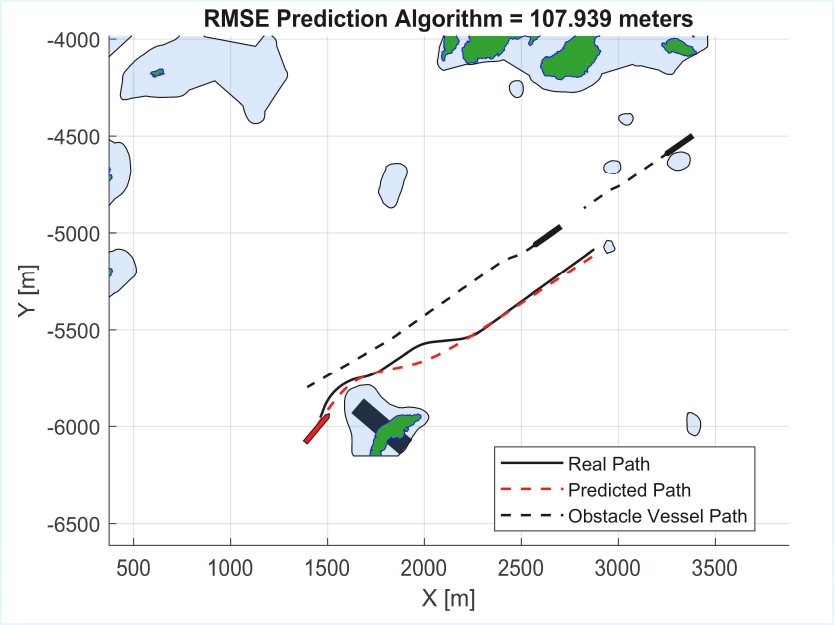}
	\caption{Prediction results for Head-On Scenario between the red target vessel and two approaching black obstacle vessels.}
	\label{DoubleHeadOn-Resultat}
\end{figure}

The first simulation is shown in Fig.~\ref{DoubleHeadOn-Resultat}, which is a recreation of an encounter in Hake Fjord outside the port of Gothenburg on the evening of August 18, 2025, where a pleasure craft faced two consecutive vessels approaching head on as it rounded a nearby island.  This is a complex scenario where the maneuver is not only dependent on the imminent obstacle ahead, but also on the following encounters with the approaching vessels afterward.
It can quickly be assessed that a CV-prediction would predict that the vessel collides with the island straight ahead which would be an unlikely outcome. Instead, the prediction algorithm handles this by identifying the appropriate COLREGs maneuvers according to Fig.~\ref{fig:colregs_sit} and utilizing the closest point of approach estimation and long-term prediction to make a realistic prediction.

\begin{figure}[htbp]
	\centering
	\includegraphics[width=0.8\linewidth]{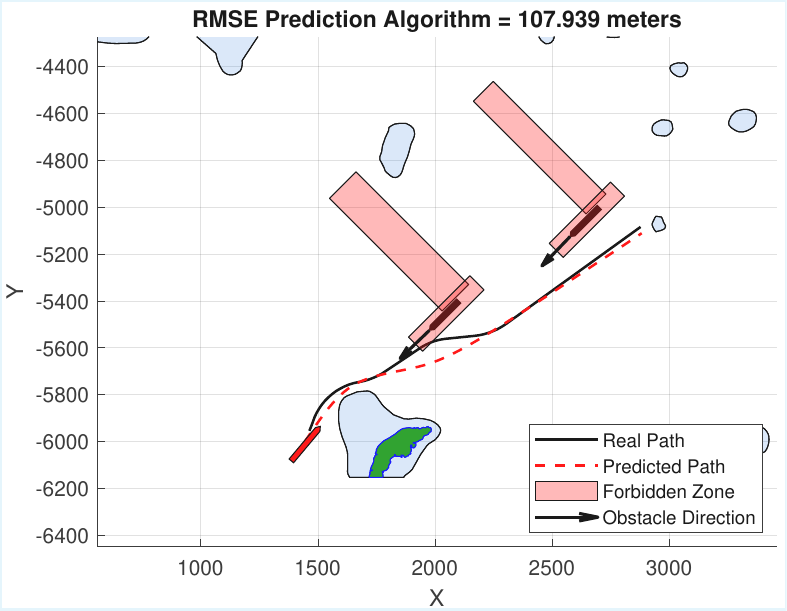}
	\caption{Proposed Prediction Algorithm interpretation of the scenario in Fig.~\ref{DoubleHeadOn-Resultat}, with generated forbidden zone and CPA-estimation.}
	\label{DoubleHeadOn-ResultatMedForbiddenzone}
\end{figure}

Fig.~\ref{DoubleHeadOn-ResultatMedForbiddenzone} shows how the proposed prediction method observes the scenario. It identifies the static obstacle straight ahead, but also considers the upcoming COLREGs scenarios. Using the results from steps 8 and 11 in Algorithm~\ref{alg:cap}, the method first estimates the CPA to identify COLREGs and generate the corresponding forbidden zone according to Fig.~\ref{fig:forbidden} for both head-on scenarios. From this, it predicts that the vessel will avoid the island, follow COLREGs, and return to its original course after the maneuver, similar to the real outcome in the scenario.
This context-aware prediction comes as a result of the long-term component in the initialization of the OCP, where a goal point has been extracted from either a goal point-estimation or AIS-data. In contrast to a CV-model, this allows the model to not only consider the feasibility of the imminent COLREGs maneuver, but also the most probable outcome once the maneuver is completed.
%This context-aware prediction considers the future movement of vessels, final destinations and nearby obstacles and manages to capture the intent of the captain without computationally heavy factors such as the impact of wind, waves and vessel dynamics. 

%This scenario highlights many weaknesses of a CV-assumption, where not only static obstacles and surrounding vessels are omitted when making a prediction, but also failing to take the final destination and long-term trajectory into account. 

%Unlike the CV-assumption, the proposed method accounts for the immediate feasibility of the upcoming COLREGs after passing the island, but also the most probable post-maneuver outcome. From this, the proposed algorithm predicts that the vessel will return to its original course after passing the obstacles, similar to the outcome of the real scenario. This context-aware prediction considers the future movements of vessels, final destinations and nearby obstacles and manages to capture the intent of the captain without computationally heavy factors such as the impact on wind, waves and vessel dynamics. 

%Looking at the position and heading of the target vessel, a CV-assumption could technically be possible. The vessel would barely clear the shallow waters around the island and cross the approaching vessels' path, ending up on their port side without colliding. However, this maneuver would still violate of the head-on COLREGs and as such it would be an improbable outcome to predict. 
\begin{figure}[htbp]
	\centering
	\includegraphics[width=0.8\linewidth]{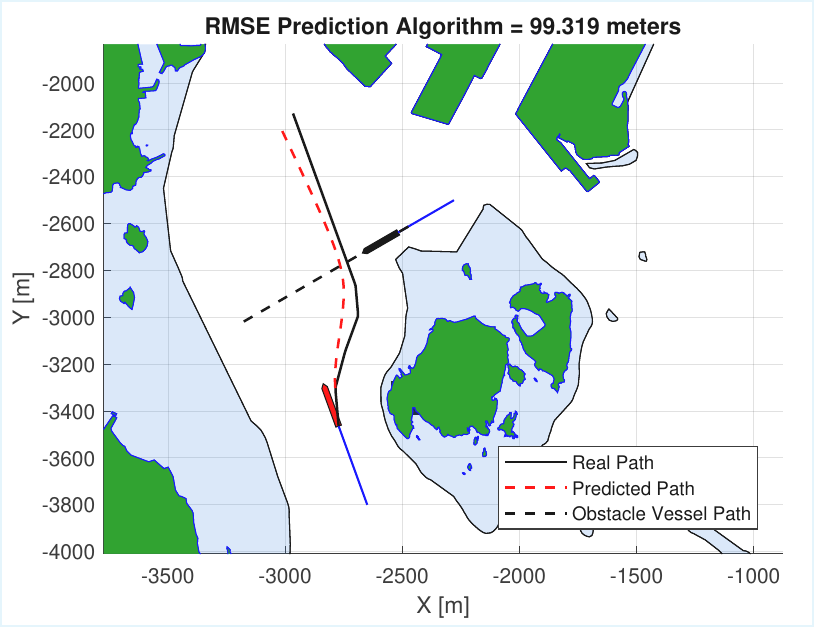}
	\caption{Prediction results for Give-Way scenario between the red target vessel and black obstacle vessel.}
	\label{GiveWay-ResultatMedObstaclepath}
\end{figure}

%\begin{figure}
%    \centering
%    \includegraphics[width=0.8\linewidth]{GiveWay-Resultat_cropped.pdf}
%\caption{GiveWay-Resultat}
%\label{GiveWay-Resultat}
%\end{figure}
Fig.~\ref{GiveWay-ResultatMedObstaclepath} shows a typical give-way scenario, captured from an encounter next to Pihlajasaari in the port of Helsinki on August 20, 2025, where the red target vessel faces an approaching vessel on its starboard side.
In this configuration, a Constant Velocity (CV) prediction performs poorly as the vessel needs to adjust its trajectory in order to comply with the give-way COLREGs, and a straight-forward prediction places the vessels on a perceived collision course.
Instead, the proposed prediction algorithm correctly identifies the give-way COLREGs and takes this into account when generating a trajectory. The predicted path is shown to have had the right intent and course of action as the path from the actual encounter in Helsinki. From the perspective of a motion planning algorithm on the black vessel, a CV-prediction may incorrectly indicate an imminent collision, prompting abrupt and unnecessary maneuvers to avoid it.
Such erratic behavior would make the entire scenario less predictable and more difficult to navigate through safely.

%\begin{figure}
%    \centering
%    \includegraphics[width=0.8\linewidth]{HeadOn-Resultat_MedForbiddenzone_cropped.pdf}
%    \caption{HeadOn-ResultatMedForbiddenzone}
%    \label{HeadOn-ResultatMedForbiddenzone}
%\end{figure}
\subsection{Impact of Prediction Errors on Motion Planning}

\begin{figure}[htbp]
	\centering
	\includegraphics[width=0.8\linewidth]{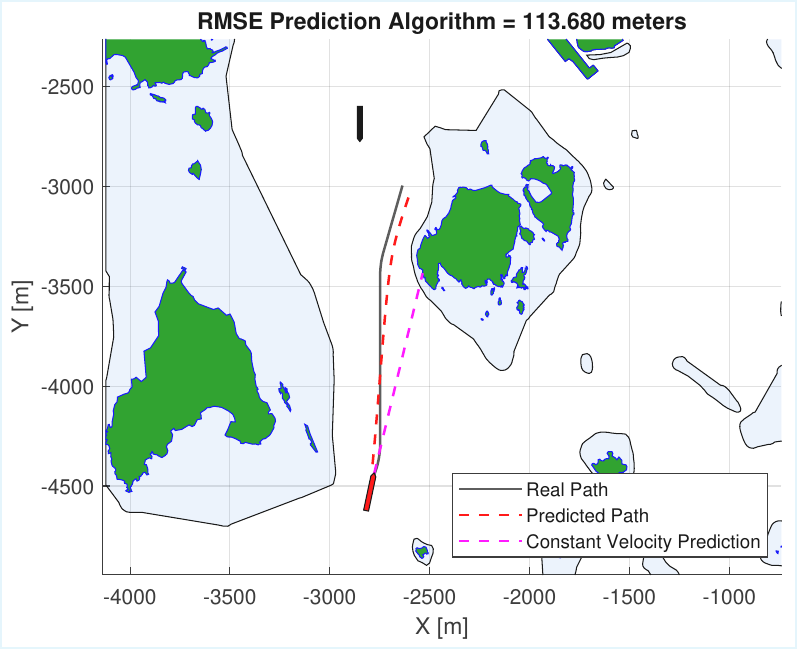}
	\caption{Prediction results for Head-On Scenario between the red target vessel and black obstacle vessel approaching from the north.}
	\label{HeadOn-ResultatMedConstantVelocity}
\end{figure}

Fig.~\ref{HeadOn-ResultatMedConstantVelocity} reflects an encounter between two vessels in the port of Helsinki on the morning of August 19, 2025. The predicted target vessel is approaching from the south with a heading that is currently straight towards the island ahead, and another vessel is approaching from the north.

This scenario shows that a prediction that does not lead to a direct collision may still indirectly create an unsafe situation by giving the planner a false sense of confidence. If the black vessel approaching from the north would plan a path based on a CV-prediction of the target vessel, it would incorrectly assume that it is going to stay on track towards the island. With this prediction, the entire narrow passage at $(x, y) = (-2700, -3500)$ would be considered safe to travel through. This showcases a limitation with CV-predictions where motion planners are given larger free spaces to navigate through than what is actually plausible. The overconfident planner risks placing the vessel in what will lead to a complex COLREGs scenario, potentially leading to an inevitable collision once the target vessel alters its course. In contrast, the proposed prediction method identifies the obstacle ahead and predicts that the target vessel will adjust its course while also considering the vessel approaching from the north.

%\begin{figure}
%\centering
%\includegraphics[width=0.8\linewidth]{HeadOn-%Resultat_MedObstaclepath_cropped.pdf}
%    \caption{HeadOn-ResultatMedObstaclepath}
%    \label{HeadOn-ResultatMedObstaclepath}
%\end{figure}

\subsection{Overtaking and Head-On Prediction}
\begin{figure}[htbp]
	\centering
	\includegraphics[width=0.8\linewidth]{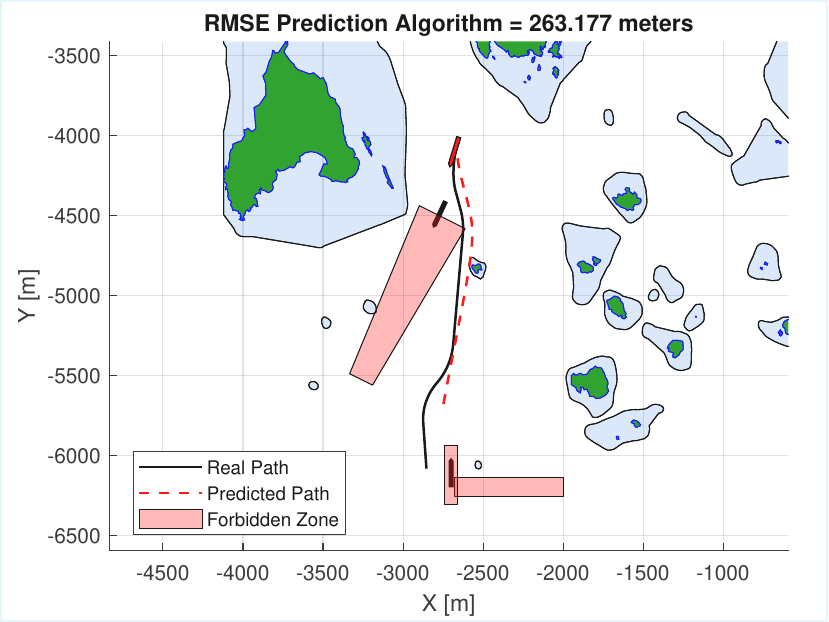}
	\caption{Predictions for Overtaking/Head-On scenario between the red target vessel and two black obstacle vessels.}
	\label{OvertakingHeadOn-ResultatmedForbiddenZones}
\end{figure}

Fig.~\ref{OvertakingHeadOn-ResultatmedForbiddenZones} shows the target vessel in red approaching from the top of the map with a southwest heading and two vessels ahead.
This encounter took place in the port of Helsinki on August 18 and is unique for this study in the sense that the vessel's current heading is not towards its final destination.
It highlights the strength of the long-term prediction in step 4 of Algorithm~\ref{alg:cap}, where the prediction towards a final destination guides the algorithm with decisions where more than one outcome may be feasible. The target vessel is allowed to overtake on either side of the vessel ahead according to Fig.~\ref{fig:colregs_sit}, but when considering the long-term prediction there is a clear bias in overtaking on the port side as it places the vessel straight towards its final destination. After the overtaking, the algorithm recognizes the approaching head-on scenario and adjusts its path similarly, but not as drastically, as the vessel in the real scenario. While predicting the exact motion of another vessel is less critical, understanding its high-level behavior, such as which side it will pass an obstacle on, is important when planning the path of the EGO ship.

There are other methods of predicting the most likely path around a vessel in an overtaking scenario. One could model complex motion dynamics for the vessels and factor in the impact of wind and waves to evaluate which maneuver is easier to execute for the captain. While this is something that the proposed method has support for in step 16 of Algorithm~\ref{alg:cap}, it comes at a computational cost and requires detailed information about the motion dynamics of the predicted vessels. Instead, the emphasis on basing a prediction on COLREGs and forbidden zones is proven to be a cheap method of adapting the prediction to the most likely actions given the scenario. Because COLREGs are expected to be followed, they provide a robust basis for the prediction method, making it adaptable even to scenarios it has not encountered before.

%\begin{figure}
%\centering
%\includegraphics[width=0.8\linewidth]{OvertakingHeadon-%Resultat_MedObstaclepath_cropped.pdf}
%    \caption{OvertakingHeadon-ResultatMedObstaclepath}
%    \label{OvertakingHeadon-ResultatMedObstaclepath}
%\end{figure}

%\begin{figure}
%\centering
%\includegraphics[width=0.8\linewidth]{OvertakingHeadon-%Resultat_MedObstaclepath_Storskala_cropped.pdf}
%\caption{OvertakingHeadon-%ResultatMedObstaclepathStorskala}
%    \label{OvertakingHeadon-ResultatMedObstaclepathStorskala}
%\end{figure}

% \subsection{Figures}

% \subsection{Tables}

% \subsection{Final Stage}

% \subsubsection{Page margins.}
% \subsection{PDF Creation}

% \subsubsection{PDFLaTeX}

\section{Conclusion}

In conclusion, this paper proposes a COLREGs-aware prediction algorithm which formulates the trajectory forecasting as an optimal control problem with the support of an approximate A*-trajectory. This is a modular approach with the ability to configure the solver similarly to how a motion planner functions. This is especially useful in previously unencountered scenarios, where a black-box approach is expected to perform poorly, by adapting the prediction by the same parameters a captain or autonomous vessel is likely to consider. Another key strength is that, unlike many machine learning and filter based methods which rely on a window of previously recorded data to make predictions, the proposed method requires only the current position, heading, and velocity. This allows it to generate a prediction more quickly when a vessel is first detected. The distance calculation to surrounding obstacles and long-term prediction is used to filter out high-risk and unfeasible scenarios that a CV-prediction often fails to identify. In the future, the algorithm could with benefit be expanded to include secondary factors affecting the navigation environment and vessel dynamics, such as wind and waves introducing forces in other degrees of freedom than those currently considered.

\bibliography{ifacconf}
% \appendix
% \section{A summary of Latin grammar}    % Each appendix must have a short title.
% \section{Some Latin vocabulary}              % Sections and subsections are supported  
% in the appendices.
\end{document}